\documentclass[showpacs]{elsart}
\usepackage{graphicx}
\usepackage{dcolumn}
\usepackage{bm}
\usepackage{amsmath}
\usepackage{amsfonts}
\usepackage{amssymb}
\usepackage{color}
\providecommand{\U}[1]{\protect\rule{.1in}{.1in}}

\begin{document}
\begin{frontmatter}
\title{Alternative thermodynamic cycle for the Stirling machine}
\author{Alejandro Romanelli},
\address{Instituto de F\'{\i}sica, Facultad de Ingenier\'{\i}a\\
Universidad de la Rep\'ublica\\ C.C. 30, C.P. 11000, Montevideo, Uruguay}
\thanks[PEPE]{Corresponding author. \textit{E-mail address:}
alejo@fing.edu.uy}
\date{\today}
\begin{abstract}
We develop an alternative thermodynamic cycle for the Stirling machine, where the polytropic process plays a
central role. Analytical expressions for pressure and temperatures of the working gas are obtained as a
function of the volume and the parameter which characterizes the polytropic process. This approach achieves a
closer agreement with the experimental pressure-volume diagram and can be adapted to any type of the Stirling
engine.
\end{abstract}
\end{frontmatter}
\maketitle
%\textbf{}
\section{Introduction}
\label{sec:I} The year $2016$ marked the bicentenary of the submission of a patent by Robert Stirling that
described his famous engine~\cite{Reid,Sier}. A Stirling engine is a mechanical device which operates in a
closed regenerative thermodynamic cycle; with cyclic compressions and expansions of the working fluid at
different temperature levels. The flow of the working fluid is controlled only by the internal volume
changes, there are no valves and there is a net conversion of heat into work or vice-versa. This engine can
run on any heat source (including solar heating), and if combustion-heated it produces very low levels of
harmful emissions. A Stirling-cycle machine can be constructed in a variety of different configurations. For
example the expansion-compression mechanisms can be embodied as turbo-machinery, as piston-cylinder, or even
using acoustic waves. Most commonly, Stirling-cycle machines use a piston-cylinder, in either an $\alpha$,
$\beta$ or $\gamma$ configuration~\cite{David}.

At the present time several researchers are working to improve the basic ideas of the Stirling
engine~\cite{CS,Senft,Kong,Barreto}. The whole engine is a sophisticated compound of simple ideas. The
challenge is to make such devices cheap enough to generate electrical power economically. It seems likely
that the biggest role for Stirling engines in the future will be to create electricity for local use.

The adequate theoretical description of the thermodynamic cycle for the Stirling machine is not obvious and
in general it is necessary to adopt certain simplifications. Usually the thermodynamical cycle is modeled by
alternating two isothermal and two isometric processes. This model has the virtue to give a simple
theoretical base, but the real thermodynamical process is quite different. In this paper we develop an
alternative approach, where the polytropic process plays a central role in the cycle. We provide an
analytical expression for the pressure of the fluid as a function of its volume and the parameter which
characterizes the polytropic process.

The paper is organized as follows.  In the next section we treat some aspects of the usual thermodynamic
cycle for the Stirling engine. In Sec.~\ref{secIII}, we introduce the alternative cycle. In Sec.~\ref{secIV},
we introduce the kinematics of the engine in order to complete the model. Finally, in Sec.~\ref{secV}, we
present the main conclusions.

\section{Usual Stirling cycle}
\label{sec:II} 
\begin{figure}[h]
\begin{center}
\includegraphics[scale=0.4]{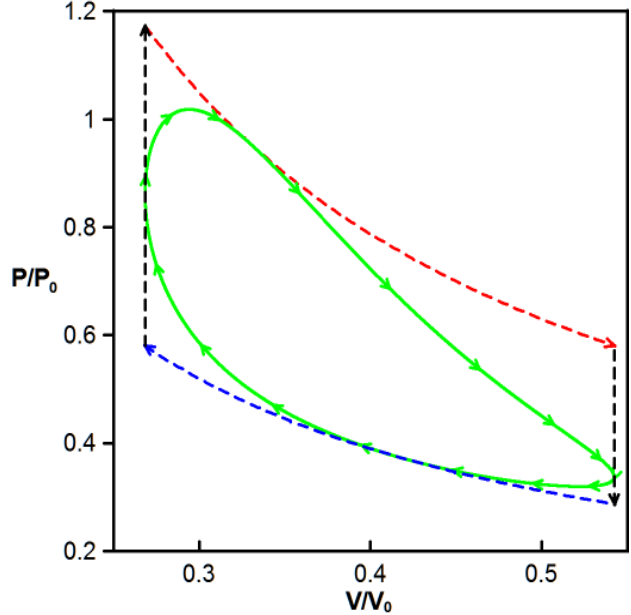}
\end{center}
\caption{Dimensionless pressure-volume diagram. The alternative Stirling cycle is presented in continuous green line
against the usual Stirling cycle in dashed line. The arrows indicate the evolution with growing $\theta$. The parameters
values are: $\phi=5.2\pi$, $\alpha=10$, $\beta=1.33$, $\varepsilon=0.1$ and $z=0.1$.} \label{f2}
\end{figure}

The usual Stirling cycle, consists of four reversible processes involving pressure and volume
changes. We present these processes plotted in dashed lines in the pressure-volume diagram of Fig.~\ref{f2}.
It is an ideal thermodynamic cycle made up of two isothermal and two isometric regenerative processes. The
relation between the movements of the pistons and the processes of the cycle are explained in basic
thermodynamics textbooks~\cite{Zeman}. The net result of the Stirling cycle is the absorption of heat $Q_H$
at the high temperature $T_H$, the rejection of heat $Q_L$ at the low temperature $T_L$, and the delivery of
work $W=Q_H+Q_L$ to the surroundings, with no net heat transfer resulting from the two constant-volume
processes. This usual cycle is very useful to understand some qualitative aspects of the Stirling machine,
however it is a coarse approach to the real experimental cycle. The thermal efficiencies of the best Stirling
engines can be as high as those of a Diesel engine~\cite{Walker}, and theoretically they have the Carnot
efficiency.
\begin{figure}[h]
\begin{center}
\includegraphics[scale=0.5]{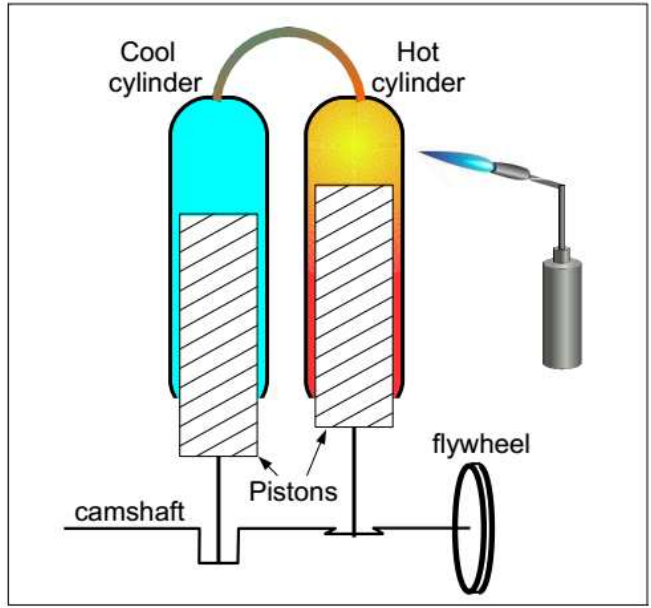}
\end{center}
\caption{Simple model of the Stirling engine. It uses two disposable glass syringes as piston-cylinder
system, see Ref.~\cite{Wagner}.} \label{f0}
\end{figure}
\begin{figure}[h]
\begin{center}
\includegraphics[scale=0.8]{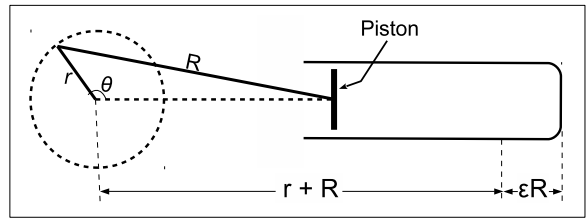}
\end{center}
\caption{Constraint scheme between the volume of gas in the cylinder and the camshaft angle $\theta$. The
circle represent the cam rotation. The piston is considered of negligible thickness.} \label{f1}
\end{figure}

Practical Stirling-cycle machines differ from the usual ideal cycle in several important aspects
~\cite{David}: (i) The regenerator and heat-exchangers in practical Stirling-cycle machines have nonzero
volume, this means that the working gas is never completely in either the hot or cold zone of the machine,
and therefore never at a uniform temperature. (ii) The motion of the pistons are usually semi-sinusoidal
rather than discontinuous, leading to non-optimal manipulation of the working gas. (iii) The expansion and
compression processes are better fitted as polytropic rather than isothermal. This allows pressure and
temperature fluctuations in the working gas and leads to adiabatic and transient heat transfer losses. (iv)
Fluid friction losses occur during gas displacement, particularly due to flow through the regenerator. (v)
Other factors such as heat conduction between the hot and cold zones of the machine, seal leakage and
friction, and friction in kinematic mechanisms all cause real Stirling-cycle machines to differ from ideal
behavior. The above factors tend to reduce the performance of real machines.
\section{Alternative Stirling cycle}
\label{secIII} In this paper we develop an alternative approach for the Stirling cycle using a simple model
for its engine. Nowadays, it easy to find in the Internet a lot of videos that present the construction and
operation of several toy models of the Stirling engine. We choose one of the simpler of these
models~\cite{Wagner} (that works very well, is not expensive and can be build by pre-university students) to
develop our ideas about its thermodynamics. This device is schematically shown in Fig.~\ref{f0}. The pistons
are connected to the camshaft with an incorporated flywheel. As the shaft rotates, the pistons move with a
constant phase difference. The two cylinders are filled with a fixed mass of air, which is recycled from one
cylinder to the other. One of the cylinders is kept in contact with the high-temperature reservoir, while the
other is in contact with the low-temperature reservoir. The connection between the cylinders is through a
small tube that may have a sponge device called regenerator. The model used in this paper has no regenerator.

As an alternative to the cycle described in the previous section, the cycle proposed in this section consists
in a polytropic process for the working gas.

Let us briefly review the main characteristics of such a process. It is a quasistatic process carried out in
such a way that the specific heat $c$ remains constant \cite{definicin1,definicin2,definicin3,romanelli}.
Therefore the relation between heat and temperature is given by
\begin{equation}
dQ=c\,N \,dT,  \label{dqdt}
\end{equation}
where $dQ$ is the heat absorbed by the gas, $T$ is the absolute temperature, and $N$ is the number of moles.
The value of~$c$ determines the relation between pressure and volume for the process. A process for which the
pressure or the volume is kept constant is, of course, polytropic with specific heat $c_P$ or $c_{V}$,
respectively. An adiabatic process is a polytropic process with $c=0$. At the other extreme, isothermal
evolution can be thought as a polytropic process with infinite specific heat. Usually textbooks define a
polytropic process of the ideal gas through
\begin{equation}
 P {V}^\beta=\text{constant},  \label{poly2}
\end{equation}
where $P$ is pressure of the gas, $V$ its volume and
\begin{equation}
\beta=\frac{c_{P}-c}{c_{V}-c}, \label{beta}
\end{equation}
is the polytropic index. It is also useful to express $c$ as a function of $\beta$
\begin{equation}
c=c_{V}\frac{\gamma-\beta}{1-\beta}.  \label{cc}
\end{equation}
where $\gamma=c_{P}/c_{V}$, and note that if $\beta\in[1,\gamma]$ then $c < 0$.

Returning to our approach, we assume some simplifications for the working gas: (i) the gas has an uniform
temperature in each cylinder, $T_2$ for the hot cylinder and $T_1$ for the cool cylinder, (ii) the mass of
gas inside the tube that connects the cylinders is negligible, (iii) due to the connection between the cold
and hot cylinders the gas has always an uniform pressure $P$, (iv) the gas is considered to behave as a
classical ideal gas. In this context, the relation between the internal energy of the gas inside the Stirling
machine and the temperature is
\begin{equation}
E=c_{V}\,\left(N_{1}\,T_{1}+N_{2}\,T_{2}\right), \label{unoa}
\end{equation}
where $c_{V}$ is constant and $N_{1}$ and $N_{2}$ are the number of moles of the gas inside of the cool and
hot cylinder respectively. From Eq.~(\ref{unoa}), the energy change is calculated to be
\begin{equation}
dE=c_{V}\,\left(N_{1}\,dT_{1}+N_{2}\,dT_{2}\right)+c_{V}\,\left(T_{1}-T_{2}\right)\,dN_{1}, \label{dosa}
\end{equation}
where we have used the conservation of the total number of moles $N=N_{1}+N_{2}$. The first term on the right
hand side of Eq.~(\ref{dosa}) takes into account the energy change associated to the change of the mean
kinetic energy of the molecules with unchanged mole numbers. The second term corresponds to the energy change
due to the mass redistribution in the cylinders, with unchanged temperatures.

In order to analyze the evolution of the system, we compare Eq.~(\ref{dosa}) with the statement of the first
law of thermodynamics
\begin{equation}
dE=dQ-dW. \label{law}
\end{equation}
The infinitesimal work is given by
\begin{equation}
dW=P\,dV, \label{re1}
\end{equation}
where $dV$ is the infinitesimal change of the total gas volume $V$.

The total absorbed (or emitted) heat $dQ$ is given by
\begin{equation}
dQ =c\,\left(N_{1}\,dT_{1}+N_{2}\,dT_{2}\right)+dQ', \label{re2}
\end{equation}
where the first term on the right hand side represents the heat associated with the polytropic process in a
bipartite system with two different temperatures, and the second term $dQ'$ is the heat associated to the gas
racking from one cylinder to the other. Therefore Eq.~(\ref{law}) may be rewritten as
\begin{equation}
dE=-P\,dV+c\,\left(N_{1}\,dT_{1}+N_{2}\,dT_{2}\right)+dQ'. \label{law1}
\end{equation}
Comparing Eq.~(\ref{dosa}) with Eq.~(\ref{law1}) and taking into account that the redistribution of the
molecules in the cylinders does not produce net work in this system it is clear that
\begin{equation}
dQ'=c_{V}\,\left(T_{1}-T_{2}\right)\,dN_{1}, \label{cuatroa}
\end{equation}
and then
\begin{equation}
PdV=(c-c_{V})\,\left(N_{1}\,dT_{1}+N_{2}\,dT_{2}\right). \label{cincoa}
\end{equation}

On the other hand, using only the equation of state of the ideal gas we obtain the following expressions for
the gas in the cylinders
\begin{equation}
\frac{N_{1}}{N}=\frac{V_{1}\,T_{2}}{V_{1}\,T_{2}+V_{2}\,T_{1}}, \label{n1}
\end{equation}
\begin{equation}
\frac{N_{2}}{N}=\frac{V_{2}\,T_{1}}{V_{1}\,T_{2}+V_{2}\,T_{1}}, \label{n2}
\end{equation}
\begin{equation}
\frac{P}{P_0}=\left(\frac{T_{1}\,T_{2}}{T_{10}\,T_{20}}\right)
\left(\frac{V_{10}\,T_{20}+V_{20}\,T_{10}}{V_{1}\,T_{2}+V_{2}\,T_{1}}\right), \label{p}
\end{equation}
where $V_{1}$ and $V_{2}$ are the volumes of gas in the cylinders, $\{V_{10}, T_{10}\}$, $\{V_{20},T_{20}\}$
and $P_0$ are the initial conditions. Note that $V_{1}+V_{2}=V$.

Using the previous results we can rewrite Eq.~(\ref{cincoa}) in the following way
\begin{equation}
V_{1}\frac{dT_{1}}{T_1}+V_{2}\frac{dT_{2}}{T_2}=(1-\beta) (dV_1+dV_2). \label{primera}
\end{equation}
Equation~(\ref{primera}) is, clearly, symmetric with respect to an interchange of the thermodynamic variables
with index $1$ and $2$. Then its solution ($T_1$ and $T_2$) must be symmetric in the arguments $V_1$ and
$V_2$ and their functional forms must be essentially the same. Therefore, it is sensible to propose the
proportionality between both temperatures $T_1$ and $T_2$, and this proposal will be verified once a solution
for Eq.~(\ref{primera}) is obtained. Then we assume
\begin{equation}
T_{2}=\alpha T_{1}, \label{alfa}
\end{equation}
where $\alpha$ is to be determined through the initial conditions
\begin{equation}
\alpha\equiv{T_{20}}/{T_{10}}. \label{alfa1}
\end{equation}
Replacing Eq.~(\ref{alfa}) in  Eq.~(\ref{primera}), it is straightforward to obtain the solutions
\begin{equation}
T_{1}=T_{10}\left(\frac{V_{0}}{V}\right)^{\beta-1},  \label{t1}
\end{equation}
\begin{equation}
T_{2}=T_{20}\left(\frac{V_{0}}{V}\right)^{\beta-1},  \label{t2}
\end{equation}
where $V_{0}$ is the total initial volume. In Appendix A we present a formal proof that Eqs. (\ref{t1}) and
~(\ref{t2}) are the solution of Eq.~(\ref{primera}).

From Eqs.~(\ref{p}),~(\ref{t1}) and ~(\ref{t2}) the pressure of the working gas is expressed as
\begin{equation}
P=P_{0}\left(\frac{\alpha\,V_{10}+V_{20}}{\alpha\,V_{1}+V_{2}}\right)\left(\frac{V_{0}}{V}\right)^{\beta-1}. \label{pe}
\end{equation}
 Equations ~(\ref{t1}),~(\ref{t2}) and ~(\ref{pe}) prove explicitly that the system undergoes a
polytropic process. Using these results and Eqs.~(\ref{re2}) and ~(\ref{cuatroa}) we obtain a differential
equation for the heat absorbed in the process,
\begin{equation}
dQ=\frac{P}{\gamma-1}\left[ (\gamma-\beta)dV+(1-\alpha) \frac{V_2dV_1-V_1dV_2}{\alpha V_{1}+V_{2}}\right].
\label{segunda}
\end{equation}
The thermodynamic properties of the Stirling engine are determined by Eqs. ~(\ref{t1}), (\ref{t2}),
(\ref{pe}) and (\ref{segunda}). These equations are the principal theoretical results of this paper. In the
next section we derive some properties of the Stirling engine, using the above results.
\begin{figure}[ht]
\begin{center}
\includegraphics[scale=0.3]{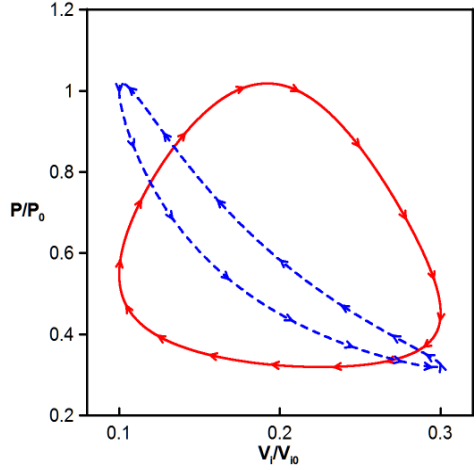}
\end{center}
\caption{Dimensionless pressure-volume diagram for the hot (continuous red line) and cool (dashed blue line)
gas. For the hot (cool) gas the abscissa axis is $V_2/V_{20}$ ($V_1/V_{10}$). The arrows indicate the
evolution with growing $\theta$. The values of the parameters are the same as in Figure ~\ref{f2}.}\label{f3}
\end{figure}
\begin{figure}[ht]
\begin{center}
\includegraphics[scale=0.4]{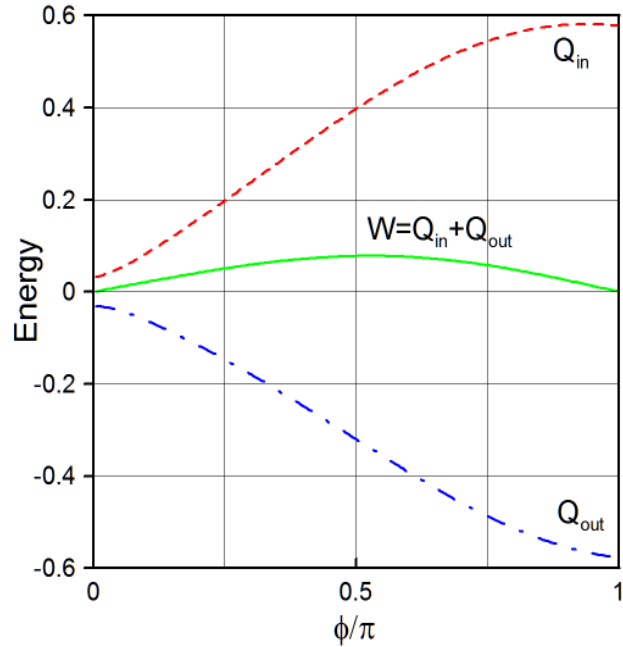}
\end{center}
\caption{The energy per cycle as a function of the phase difference between the cams $\phi$. Solid green line
the work, dashed red line the heat absorbed, dashed-dot blue line the heat rejected. The values of the
parameters are the same as in Figure ~\ref{f2}. The maximum work corresponds to $\phi=0.52\pi $.} \label{f4}
\end{figure}
\begin{figure}[ht]
\begin{center}
\includegraphics[scale=0.3]{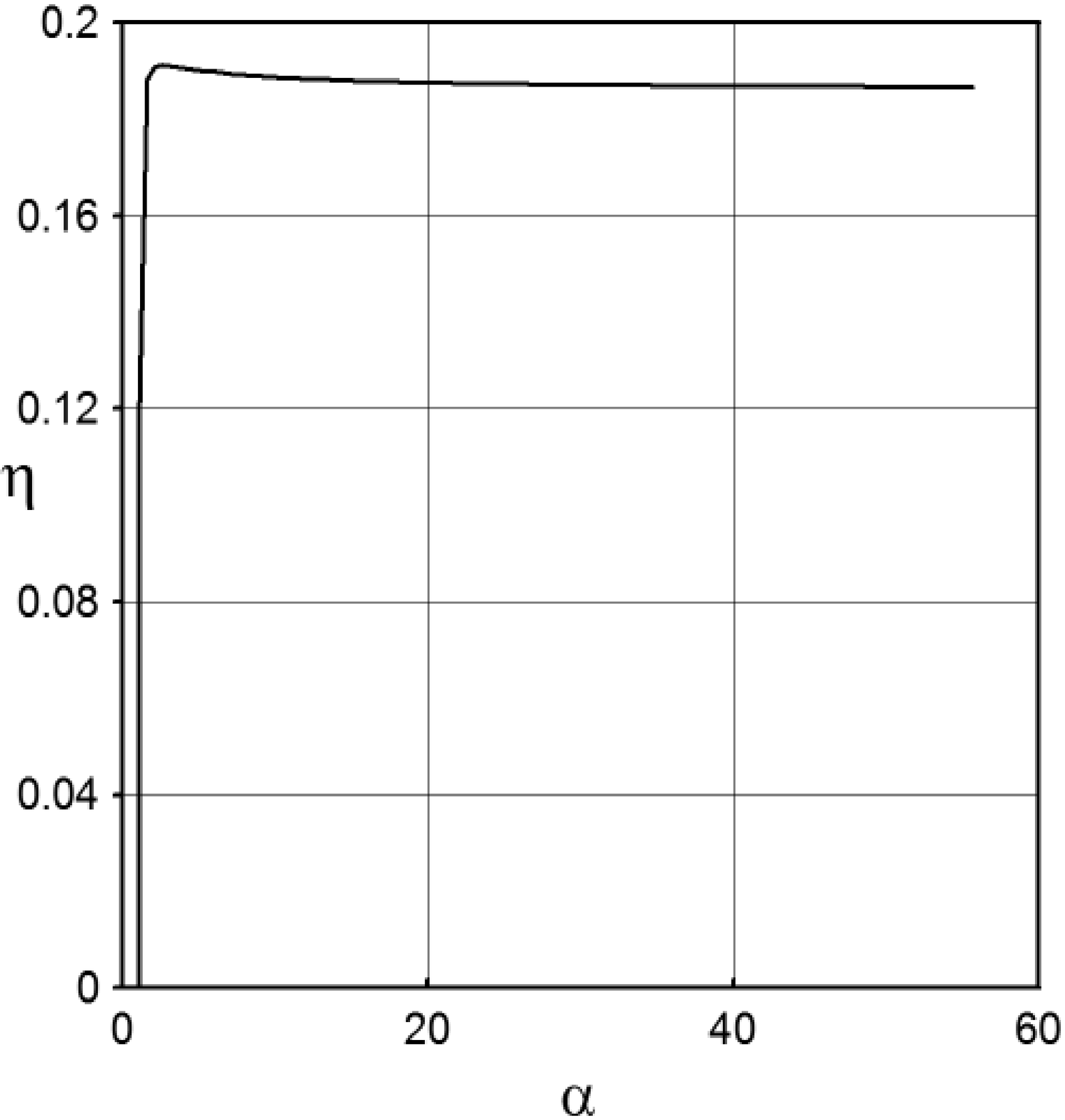}
\end{center}
\caption{The efficiency as a function of the parameter $\alpha$, with $\phi=0.52\pi$, $\beta=1.33$,
$\varepsilon=0.1$ and $z=0.1$.} \label{f5}
\end{figure}
\section{Numerical implementation}
\label{secIV}

In order to complete the description of the Stirling engine we develop a simple model for the interaction
between the pistons of the engine and the working gas. With this model it is possible to implement numerical
calculations and to obtain some characteristic results of the Stirling engine.

The operation of the Stirling engine can be described essentially in the following way. The working gas
follows the thermodynamic cycle interacting with the pistons, which are connected with the camshaft and
perform periodic movements associated to the flywheel rotation. The inertia of the flywheel collaborates to
sustain the continuity of the movement.

Figure~\ref{f1} shows the details of the connection between a piston and the camshaft through the rod of length $R$. The
cam has a rotating radius $r$ and the piston has a section of area $a$. From this figure, the functional relation between
the gas volume and the camshaft angle, $\theta$, for both cylinders, is obtained
\begin{equation}
\frac{V_1}{R\,a}=1+\varepsilon+z\left(1-\cos\theta\right)-\sqrt{1-z^{2}\cos^{2}\theta}, \label{v1}
\end{equation}
\begin{equation}
\frac{V_2}{R\,a}=1+\varepsilon+z\left[1-\cos(\theta+\phi)\right]-\sqrt{1-z^{2}\cos^{2}(\theta+\phi)},
\label{v2}
\end{equation}
where $z={r}/{R}$, $\varepsilon$ determines the minimum volume of gas in each cylinder during the cycle (we
assume the same value for both cylinders) and $\phi$ is the phase difference between the cams. In what
follows we obtain the initial conditions using $\theta_0=0$.

The pressure-volume diagram for the alternative cycle is shown in Fig.~\ref{f2}, which is obtained using
Eqs.~(\ref{pe}), (\ref{v1}) and (\ref{v2}) and additional numerical calculation. We obtain a smooth
continuous curve closer to the experimental behavior. The usual cycle is also shown for comparison with the
same highest and lowest temperatures than the alternative cycle. The area inside the continuous green line
represents the total work of the cycle;
\begin{equation}
W=\oint P\,dV=\int_{0}^{2\pi}P\,\frac{dV}{d\theta}\,d\theta, \label{work}
\end{equation}
this is the work available for overcoming mechanical friction losses and for providing useful power to the
engine crankshaft.

The alternative cycle has not the four processes of the ideal cycle sharply defined. This approach allows to
model adequately the following facts. (i) The compression and expansion processes do not take place wholly in
one or other of the cylinders. (ii) The motion of the pistons are continuous rather than discontinuous. (iii)
The heat exchange between gas and environment is best modeled by a polytropic process rather than an
isothermal-isometric sequence.

The pressure-volume diagrams for the hot and cool cylinders are shown in Fig.~\ref{f3}. They are obtained
numerically using again the same data of Fig.~\ref{f2}. As in the previous figure the areas of the curves
represent the work of each cycle. It is seen that their orientations are opposite, then the available work
for the engine is proportional to the difference between these areas.

In Fig.~\ref{f4} the work per cycle of the Stirling machine is presented as a function of $\phi$. The work
has a maximum value which depends on the initial conditions and the parameters, as given in the caption. It
is known empirically that to obtain the maximum power of the machine the phase difference between the two
cams must be near $\pi/2$. The present model gives us the precise angle to obtain numerically this maximum.

Let us call $Q_{in}$ and $Q_{out}$ the heat absorbed and rejected by the gas in the cycle. They are
calculated numerically using Eq.~(\ref{segunda}) with the convention $Q_{in}>0$ and $Q_{out}<0$, that is
\begin{equation}
Q_{in}=\int_{\mathcal{C}{in}}\frac{dQ}{d\theta}\,d\theta, \label{c1}
\end{equation}
\begin{equation}
Q_{out}=\int_{\mathcal{C}{out}}\frac{dQ}{d\theta}\,d\theta, \label{c2}
\end{equation}
where $\mathcal{C}{in}$ and $\mathcal{C}{out}$ refer to the paths where $dQ/\,d\theta>0$ and $dQ/\,d\theta<0$
respectively. As the internal energy change in the entire cycle vanishes, then the total work verifies
$W=Q_{in}+Q_{out}$ as shown in Fig.~\ref{f4}.
\begin{figure}[ht]
\begin{center}
\includegraphics[scale=0.3]{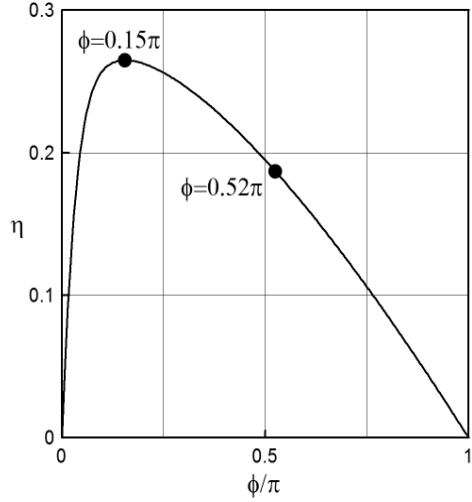}
\end{center}
\caption{The efficiency as a function of $\phi$, with $\alpha=10$, $\beta=1.33$, $\varepsilon=0.1$ and
$z=0.1$. The dots in the curve indicate the values of $\phi$ for the maximum efficiency and maximum work.}
\label{f6}
\end{figure}
\begin{figure}[ht]
\begin{center}
\includegraphics[scale=0.3]{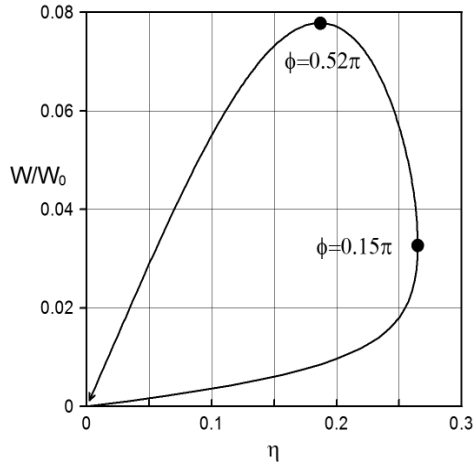}
\end{center}
\caption{Dimensionless work as a function of the efficiency, the angle $\phi$ varying between $0$ and $\pi$.
The arrow in the curve indicates the growing of $\phi$. The parameters are $W_0=P_0V_0$, $\alpha=10$,
$\beta=1.33$, $\varepsilon=0.1$, $z=0.1$.} \label{f8}
\end{figure}
\begin{figure}[ht]
\begin{center}
\includegraphics[scale=0.3]{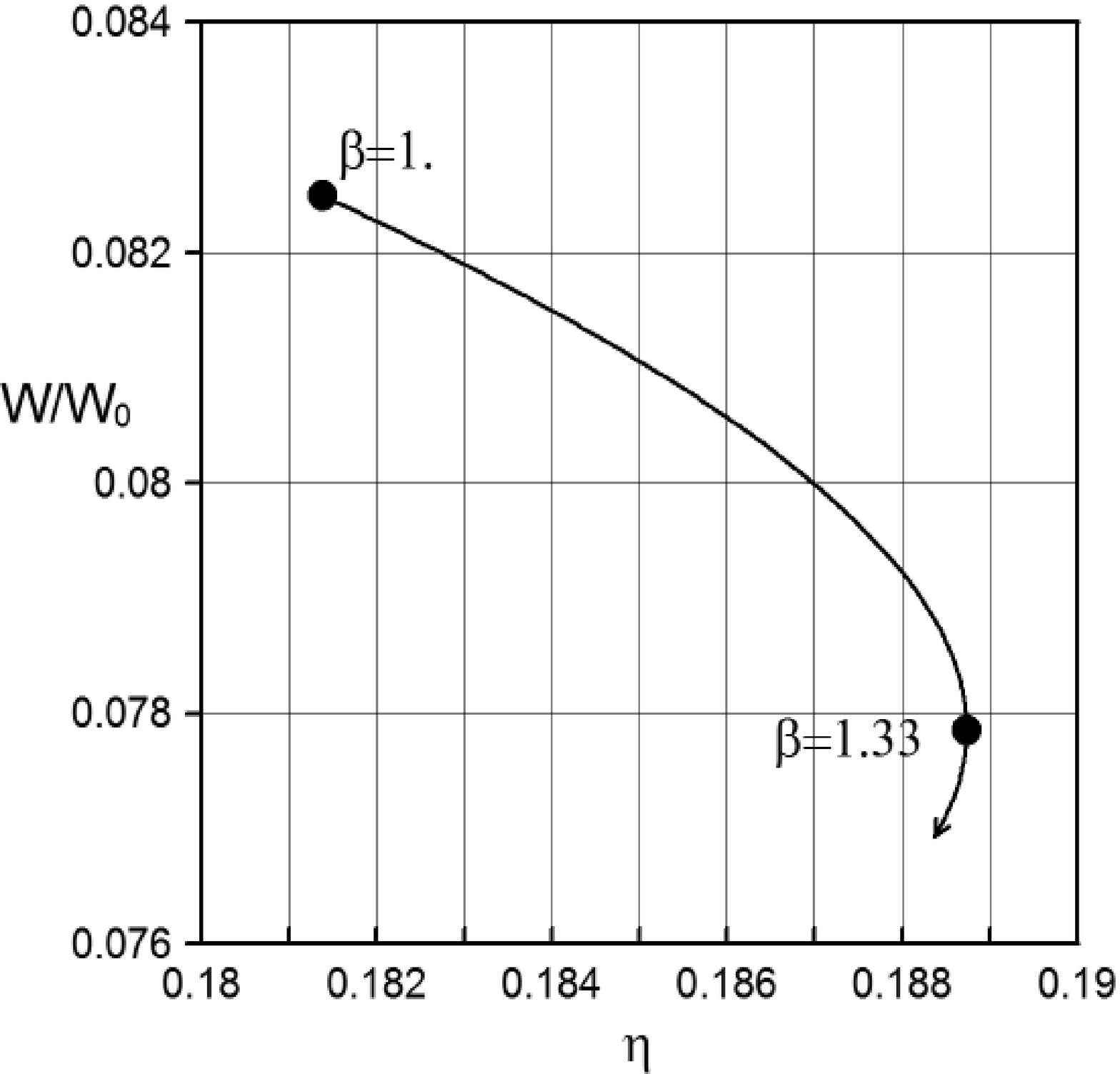}
\end{center}
\caption{Dimensionless work as a function of the efficiency, with $W_0=P_0V_0$, $\alpha=10$,
$\varepsilon=0.1$, $\phi=0.52\pi$ and $z=0.1$. The polytropic index varies from $\beta=1.0$ to $\beta=1.4$.
The arrow in the curve indicates the growing of $\beta$ } \label{f9}
\end{figure}

The machine efficiency is defined as $W/Q_{in}$, or equivalently
\begin{equation}
\eta=1+{Q_{out}}/{Q_{in}}. \label{eta}
\end{equation}
The efficiency as a function of $\alpha$ is shown in Figure~\ref{f5}. It is easy to prove from
Eqs.~(\ref{pe}) and (\ref{segunda}) that if $\alpha=1$ then $\eta=0$, this means, as expected, that without a
difference of temperature the machine does not work. When $\alpha$ grows $\eta$ also grows, however the
efficiency is asymptotically bounded by some value below $0.2$. Additionally, we have numerically checked
that the efficiency can be improved notoriously reducing the value of $\varepsilon$.

Figure~\ref{f6} shows the efficiency as a function of the phase difference between the cams. Note that the
maximum efficiency does not coincide with the maximum work, this fact is highlighted in Fig.~\ref{f8} that
shows directly the dependence between work and efficiency with varying $\phi$.

Figure~\ref{f9} shows the behavior of the work and the efficiency when the parameter $\beta$ varies. Here we
point out that in ref. ~\cite{romanelli} an analytical expression for the polytropic index of air was
obtained showing that $\beta$ depends only on the thermodynamic initial conditions.

Finally, it is interesting to underline that our results given by Eqs.~(\ref{t1},\ref{t2},\ref{pe}) for the
isothermal case $\beta=1$ ($c=-\infty$) coincide with the well-known Schmidt
solution~\cite{Schmidt,Bercho,Formosa}, published in the year $1871$.
%Therefore, in the frame of the present work, the thermodynamic solution proposed by Schmidt respond to a one
%polytropic process with an infinite specific heat.
\section{Conclusions}
\label{secV} The paper develops an alternative theoretical approach to the usual Stirling thermodynamic
cycle. This alternative cycle is obtained applying the first law of thermodynamics to the working fluid
inside the Stirling engine. The main characteristic of this approach is the introduction of a polytropic
process as a way to represent the exchange of heat with the environment.

We have obtained analytical expressions for the pressure, temperatures, work and heat for the gas inside the
engine. The theoretical pressure-volume diagram, shows a qualitative agreement with the experimental diagram.
With the aim to complete the description of the Stirling engine we develop a simple model for the interaction
between the pistons of the engine and the working gas and we study the power and the efficiency of the
Stirling engine as a function of: (i) the phase difference between the cams of the engine crankshaft, $\phi$;
(ii) the ratio of the temperatures of the heat sources, $\alpha$; (iii) the type of polytropic process,
$\beta$.

We emphasize that the Schmidt analysis for the Stirling machine is a particular case of the solution presented in this
paper for $\beta=1$.

In summary, the theoretical approach proposed in this paper describes the thermodynamics of the Stirling
engine in a simple, precise and natural way that can be adapted to any variant of this engine.

\section*{Acknowledgements}
I acknowledge the stimulating discussions with V\'{\i}ctor Micenmacher, Silvia Cedr\'ez, Italo Bove, Pedro
Curto and the support from ANII and PEDECIBA (Uruguay).

\appendix
\section*{Appendix A}
Equation~(\ref{primera}) can also be written in the symmetrical form
\begin{equation}
\left(\frac{V_{1}}{T_1}\frac{\partial{T_{1}}}{\partial{V_{1}}}+\frac{V_{2}}{T_2}\frac{\partial{T_{2}}}{\partial{V_{1}}}
-1+\beta\right)dV_{1}+\left(\frac{V_{1}}{T_1}\frac{\partial{T_{1}}}{\partial{V_{2}}}+\frac{V_{2}}{T_2}
\frac{\partial{T_{2}}}{\partial{V_{2}}} -1+\beta\right)dV_{2}=0, \notag \eqno(A.1)
\end{equation}
where we have introduced the partial differentiation because it is clear that $T_1$ and $T_2$ must depend on
both volumes $V_1$ and $V_2$. In this equation the variations $dV_{1}$ and $dV_{2}$ are completely arbitrary.
Accordingly, the only way to satisfy this condition is that both expressions between parentheses of
Eq.~(\emph{A}.1) must vanish. Then it is easy to show that Eqs.~(\ref{t1}) and ~(\ref{t2}) satisfy both
requisites.


\begin{thebibliography}{10}
\bibitem{Reid} J.S. Reid, arXiv:1604.02362 (2016).

\bibitem{Sier} R.Sier, \emph{Hot Air Caloric and Stirling Engines: A History}, Vol.1 (1999).

\bibitem{David} D.Haywood, \emph{An introduction to Stirling-cycle machines}, Stirling-cycle Research Group,
Department of Mechanical Engineering University of Canterbury (2006).

\bibitem{CS} S.C.Costa, M.Tutar, I.Barreno, J.A.Esnaola, H.Barrutia, D.García,
M.A.Gonzalez, J.I.Prieto, {Energy Convers. Manage.},
 \textbf{72}, 800 (2014).\\S.C.Costa, H.Barrutia, J.A.Esnaola, M.Tutar, {Energy Convers. Manage.},
 \textbf{79}, 225 (2014).
%Centro Stirling,\\ \emph{http://www.centrostirling.com/ing/index.html}, (2009).

\bibitem{Senft} J.R.Senft, Int. J. Energy Res., \textbf{22}, 991 (1998)

\bibitem{Kong} B.Kongtragool, S.Wongwises, Renewable Energy, \textbf{31}, 345 (2006)

\bibitem{Barreto} G.Barreto, P.Canhoto, {Energy Convers. Manage.} \textbf{132}, 119 (2017).

\bibitem{Zeman} M.W.Zemansky, R.H.Dittman, \emph{Heat and Thermodynamics},(McGraw-Hill, London, 1997).

\bibitem{Walker} G.Walker, J.R.Senft Free, \emph{Free Piston Stirling Engines}, (Springer-Verlag,
Berlin,1980).

\bibitem{Wagner}L.Wagner, in: \\ \emph{https://www.youtube.com/watch?v=dEIQxu6aU4g} (2013).

\bibitem{definicin1} G.P.Horedt, \emph{Polytropes: Applications in
Astrophysics and Related Fields}, (Kluwer, London, 2004).

\bibitem{definicin2} R.P.Drake, \emph{High-Energy-Density Physics:
Fundamentals, Inertial Fusion, and Experimental Astrophysics}, (Springer-Verlag, Berlin, 2006).

\bibitem{definicin3} S.Chandrasekhar, \emph{An Introduction to the Study of
Stellar Structure}, (Dover, New York, 1967).

\bibitem{romanelli} A.Romanelli, I.Bove, F.Gonz\'alez, \emph{Air expansion in a water rocket},
 Am. J. Phys. \textbf{81}, 762 (2013); doi: 10.1119/1.4811116

\bibitem{Schmidt} G.Schmidt, \emph{Theorie der Lehmann'schen kalorischen Maschine},
 Z Vereines Deutscher Ingenieure \textbf{15}, 1 (1871).

\bibitem{Bercho} I.Urieli, D.Berchowitz, \emph{Stirling Cycle Engine Analysis}, (International Public Service, 1983).

\bibitem{Formosa} F.Formosa, G.Despesse, {Energy Convers. Manage.}, \textbf{51}, 1855 (2010).

\end{thebibliography}
\end{document}